\documentclass{ws-procs975x65}
\def\trimmarks{}

\begin{document}

\title{Binary-pulsar tests of strong-field gravity\\
and gravitational radiation damping\footnote{\uppercase{C}ontribution
to tenth \uppercase{M}arcel \uppercase{G}rossmann \uppercase{M}eeting,
20--26 July 2003, \uppercase{R}io de \uppercase{J}aneiro,
\uppercase{B}razil}}

\author{Gilles ESPOSITO-FARESE}

\address{$\mathcal{G}\mathbb{R}\varepsilon\mathbb{C}\mathcal{O}$,
FRE 2435-CNRS, Institut d'Astrophysique de Paris,\\
98bis boulevard Arago, F-75014
Paris, France\\
E-mail: gef@iap.fr}

\maketitle

\abstracts{This talk reviews the constraints imposed by binary-pulsar
data on gravity theories, focusing on ``tensor-scalar'' ones which are
the best motivated alternatives to general relativity. We recall that
binary-pulsar tests are qualitatively different from solar-system
experiments, because of nonperturbative strong-field effects which can
occur in compact objects like neutron stars, and because one can
observe the effect of gravitational radiation damping. Some theories
which are strictly indistinguishable from general relativity in the
solar system are ruled out by binary-pulsar observations. During the
last months, several impressive new experimental data have been
published. Today, the most constraining binary pulsar is no longer the
celebrated (Hulse-Taylor) PSR B1913$+$16, but the neutron star-white
dwarf system PSR J1141$-$6545. In particular, in a region of the
``theory space'', solar-system tests were known to give the tightest
constraints; PSR J1141$-$6545 is now almost as powerful. We also
comment on the possible scalar-field effects for the detection of
gravitational waves with future interferometers. The presence of a
scalar partner to the graviton might be detectable with the LISA space
experiment, but we already know that it would have a negligible effect
for LIGO and VIRGO, so that the general relativistic wave templates can
be used securely for these ground interferometers.}

\section{Introduction and solar-system constraints}
The most efficient way to test a theory is to contrast its predictions
with alternative models. Instead of just confirming or ruling out a
particular theory, this method allows us to understand what features
have actually been tested, and what kind of observations could be
performed to test the remaining features.

The best known example of such an embedding of general relativity in a
space of alternatives is the so-called ``parametrized post-Newtonian''
(PPN) formalism,\cite{w93} which describes all possible metric
theories of gravity in weak-field conditions, at order $1/c^2$ with
respect to the Newtonian interaction. The basic idea was formulated
by Eddington,\cite{edd} who wrote the usual Schwarzschild metric in
isotropic coordinates, but introduced some phenomenological parameters
$\beta^\text{PPN}$ and $\gamma^\text{PPN}$ in front of the
different powers of the dimensionless ratio $Gm/rc^2$:
\begin{subequations}
\label{1}
\begin{eqnarray}
-g_{00}&=& 1 - 2\frac{Gm}{rc^2} + 2 \beta^\text{PPN}
\left(\frac{Gm}{rc^2}\right)^2 +
\mathcal{O}\left(\frac{1}{c^6}\right)\,,
\label{1a}\\
g_{ij}&=&\delta_{ij}\left(1+2\gamma^\text{PPN}\frac{Gm}{rc^2}\right)
+ \mathcal{O}\left(\frac{1}{c^4}\right)\,.
\label{1b}
\end{eqnarray}
\end{subequations}
General relativity corresponds to $\beta^\text{PPN} = \gamma^\text{PPN}
= 1$, and is in perfect agreement with solar system experiments. At the
time of the 10th Marcel Grossmann Meeting (MGX), the tightest published
bounds on these parameters were\cite{w01}
\begin{equation}
|\gamma^\text{PPN}-1| < 2\times 10^{-3}\,, \qquad
|\beta^\text{PPN}-1| < 6\times 10^{-4}\,.
\label{2}
\end{equation}
An unpublished\cite{eubanks} stronger constraint was also known,
$|\gamma^\text{PPN}-1| < 4\times 10^{-4}$, but an impressive new
result\cite{cassini} has been released two months after MGX:
\begin{equation}
\gamma^\text{PPN} - 1 = (2.1\pm2.3)\times 10^{-5}\,.
\label{3}
\end{equation}
Such bounds tell us that general relativity is basically the only
theory consistent with experiment at the first post-Newtonian order.
However, they do not constrain higher order terms in metric~(\ref{1}),
and the correct theory of gravity might differ significantly from
general relativity in strong field conditions. Indeed, if $R$ denotes
the radius of a body, the ratio $Gm/Rc^2 \approx 10^{-9}$ for the Earth
and $\approx 10^{-6}$ for the Sun, but it reaches $\approx 0.2$ for
neutron stars, not far from the theoretical maximum of $\frac{1}{2}$
for black holes. Pulsar observations can thus be used to test the
strong-field regime of gravity.

\section{Binary-pulsar tests}
Experiment tells us that isolated pulsars are very stable clocks. A
binary pulsar is thus a moving clock, the best tool that one could
dream of to test a relativistic theory. Indeed, as the pulsar moves
around its companion, the longitudinal Doppler effect modifies its
observed spin rate, and we get a stroboscopic information on its
orbital velocity. Several parameters characterizing its Keplerian orbit
can then be extracted from an analysis of the pulse arrival times. For
instance, the time between two maxima of the pulse frequency gives a
measure of the orbital period $P$. One can also extract the
eccentricity $e$, the longitude of periastron $\omega$, and the
projected semimajor axis $x$ along the line of sight ($x \equiv
\frac{a}{c} \sin i$, where $i$ denotes the inclination of the orbit
with respect to the plane of the sky).

\begin{figure}[ht]
\centerline{\epsfbox{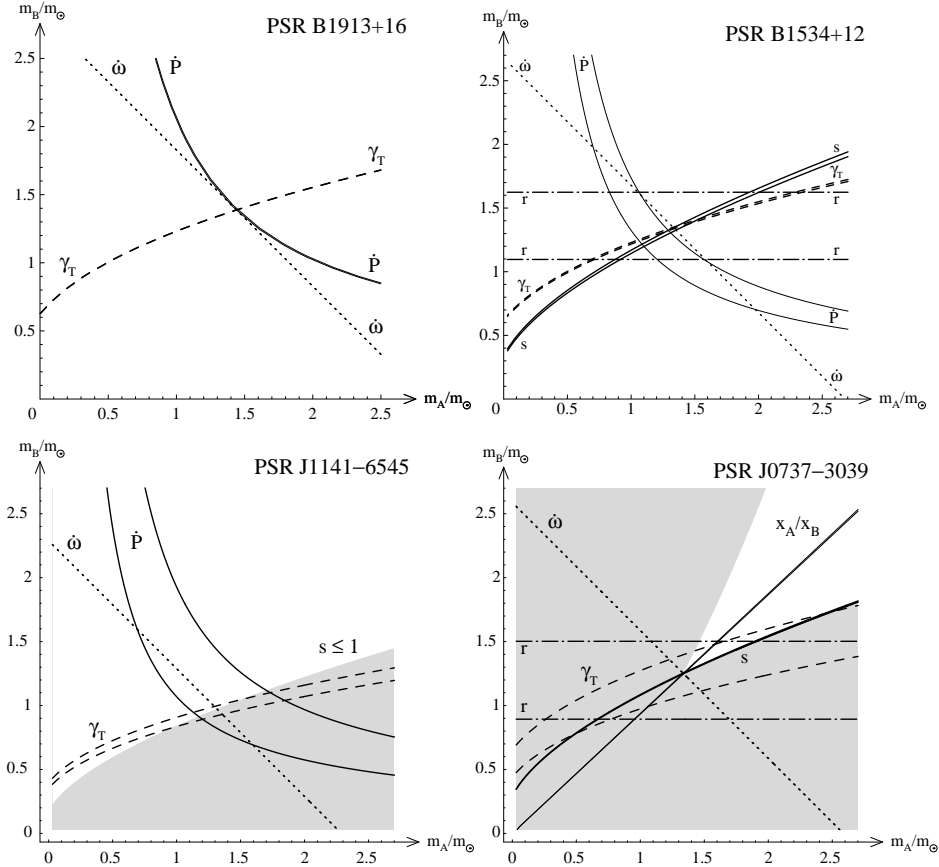}}
\caption{Mass plane ($m_A=$ pulsar, $m_B$ = companion) in general
relativity for the four precisely timed binary pulsars. The various
strips are consistent at the 1$\sigma$ level with the timing
observables which label them. In the lower two plots, the shaded region
corresponds to $|\sin i| > 1$ and is thus excluded.\label{fig1}}
\end{figure}

If high enough precision is achieved, relativistic corrections can be
measured. In particular, if a system is observed long enough, the time
derivatives of the Keplerian parameters become available. In the case
of the famous Hulse-Taylor binary pulsar \textbf{PSR B1913+16}, three
such ``post-Keplerian'' observables have been determined with great
accuracy:\cite{1913} (i)~the Einstein time delay parameter $\gamma_T$,
which combines the second-order Doppler effect ($\propto v_A^2/2 c^2$,
where $v_A$ is the pulsar's velocity) together with the redshift due to
the companion ($\propto G m_B/r_{AB} c^2$, where $m_B$ is the
companion's mass and $r_{AB}$ the pulsar-companion distance); (ii)~the
periastron advance $\dot\omega$ (relativistic effect of order
$v^2/c^2$); and (iii)~the rate of change of the orbital period, $\dot
P$, caused by gravitational radiation damping (an effect of order
$v^5/c^5$ in general relativity, but generically of order $v^3/c^3$ in
alternative theories, see below). In any theory of gravity, these three
quantities can be computed in terms of the two unknown masses of the
pulsar and its companion, and the equations
$\textit{predictions}(m_A,m_B) = \textit{observed values}$ define
three curves in the mass plane (or rather three strips if one takes
into account experimental uncertainties). This provides $3
\text{ (observables)}-2 \text{ (unknown masses)} = 1$ test of the
theory. If the three strips meet in a small region, there exists a pair
of masses consistent with all three observables, and the theory passes
the test. If they do not meet, the theory is ruled out. The upper-left
panel of Fig.~\ref{fig1} displays this mass plane in the case of
general relativity, which is nicely consistent with experimental data
within $1\sigma$ error bars. This Figure also displays the mass plane
for three other binary pulsars which have been timed accurately.

For \textbf{PSR B1534+12}, five post-Keplerian observables have been
measured:\cite{1534} $\gamma_T$, $\dot\omega$ and $\dot P$ have been
described above, whereas $r=G m_B/c^3$ and $s=\sin i$ denote
respectively the ``range'' (global factor) and the ``shape'' of the
Shapiro time delay. General relativity passes again the $5-2=3$ tests
(at the $1.3\sigma$ level for $\dot P$, whose measurement is spoiled by
a poorly known Doppler contribution due to the acceleration of the
system towards the center of the Galaxy).

The timing measurements of $\gamma_T$, $\dot\omega$ and $\dot P$ for
\textbf{PSR J1141$-$6545} are brand-new,\cite{1141} and have been
presented at this parallel session PT1 of the MGX meeting (see
M.~Bailes' contribution to the present proceedings). This binary pulsar
was discovered in 1999, and has been quickly recognized as a
``strange'' system, which triggered several detailed studies. The
pulsar is young ($\sim 1.4$ Myr) and thereby nonrecycled, as indicated
by its rather slow pulse period ($0.4$ s). The orbital period is short
($P = 4$h $45$min), therefore large relativistic effects are expected.
The companion is a white dwarf at the 90\% confidence
level,\cite{1141form} but the eccentricity of the orbit ($e = 0.17$) is
surprisingly large of such an asymmetrical system. Indeed, almost all
known neutron star-white dwarf binaries have a vanishingly small
eccentricity. The explanation is that the neutron star was formed
\textit{after} the white dwarf in PSR J1141$-$6545:\cite{1141form}
Initially, the progenitor's mass was too small to evolve into a neutron
star, but it accreted enough matter from its companion, and it finally
exploded as a type Ib/c supernova, giving a momentum kick to the
newborn neutron star. Before precise timing measurements could be
performed, the Keplerian parameters of this system could be determined
by an analysis of the pulsar scintillation.\cite{1141scint} This
phenomenon, caused by diffraction in the interstellar medium, happens
to have a characteristic timescale of minutes for PSR J1141$-$6545, and
this is quick enough to extract orbital information. Indeed, the
periodic variation of this scintillation timescale can be interpreted
as a consequence of the orbital motion of the pulsar. A fit of the
observational data gave two possible orbital solutions,\cite{1141scint}
only one of them predicting a pulsar mass $\approx 1.3 m_\odot$
consistent with all other known neutron star masses $(1.35 \pm 0.04)
m_\odot$. [The fact that it is the lightest known is consistent with
the formation scenario summarized above.] The recent timing
measurements\cite{1141} of this system nicely confirm (within general
relativity) this orbital solution, as shown in the lower-left panel of
Fig.~\ref{fig1}. Indeed, the intersection of the $\dot\omega$ and
$\gamma_T$ curves gives for the masses $m_A/m_\odot = 1.30\pm 0.02$ and
$m_B/m_\odot = 0.986\pm 0.02$. Kepler's third law implies the following
relation between the inclination angle $i$, the masses, and the
observed quantities $P$ and $x$:
\begin{equation}
\frac{(m_B \sin i)^3}{(m_A+m_B)^2} =
\left(\frac{2\pi}{P}\right)^2\frac{(xc)^3}{G}\,.
\label{4}
\end{equation}
Therefore, once $m_A$ and $m_B$ have been determined thanks to two
precise enough post-Keplerian observables, this ``mass function''
(\ref{4}) provides the inclination angle of the orbit. One finds $i >
75^\circ$ at the $1\sigma$ level, showing that the orbit in nearly
edge on. In such a situation, the mathematical bound\footnote{Note that
the Shapiro post-Keplerian parameter $s = \sin i$ has not yet been
measured precisely enough to provide another test with PSR
J1141$-$6545 (it should be available within a few months). We are here
using only the mathematical fact that it cannot be greater than 1.}
$|\sin i| \leq 1$ can be used as an extra constraint in the mass plane
$(m_A,m_B)$. The excluded region is shaded in Fig.~\ref{fig1}, and we
do verify that the intersection of the strips is very close to the
limit $\sin i = 1$. This extra constraint will be quite useful
below to exclude some alternative theories of gravity. For the above
masses and the observed Keplerian parameters of the orbit, general
relativity predicts an orbital period derivative $\dot P = -3.8\times
10^{-13}$, nicely consistent with the observed value $\dot P^\text{obs}
= (-4\pm 1)\times 10^{-13}$, as illustrated also in the lower-left
panel of Fig.~\ref{fig1}. The $1\sigma$ relative errors are still
rather large, but since they scale as $t^{-5/2}$, they should reach
$\sim 1\%$ by 2010. We will see anyway, in Sec.~4 below, that the
present precision is already extremely constraining for alternative
theories, and that PSR J1141$-$6545 is the best system available to
test gravity in the strong-field regime.

The timing measurements for \textbf{PSR J0737$-$3039} were
released\cite{0737a,0737b} five months after the MGX meeting, but this
system is so interesting that I must mention it here. Like PSRs
B1913+16 and B1534+12, this is a double neutron star system, but its
orbital period is so short ($P = 2$h $27$min) that huge relativistic
effects are expected. [This short orbital period also implies that the
system will merge in about $85$ Myr, much quicker than any other known
neutron star binary; this suffices to increase by a factor 10 the
estimated merger rate in our Galaxy,\cite{0737a} and thereby our
chances to observe a strong gravitational wave signal in the LIGO/VIRGO
interferometers.] The periastron advance could be determined in only a
few days of observation, and its value $\dot\omega =
17^\circ/\text{yr}$ is much greater than any of the above binary
pulsars (the largest being $5.3^\circ/\text{yr}$ for PSR J1141$-$6545).
A careful analysis of the data taken during several months also
provided three other post-Keplerian parameters,\cite{0737b}
corresponding to the Einstein ($\gamma_T$) and the Shapiro ($r$ and
$s$) time delays. The measure of $s = \sin i$ shows that the orbit is
again almost perfectly edge on: $i = (87 \pm 3)^\circ$. Therefore, the
mathematical bound $|\sin i| \leq 1$ can again be used as an extra
constraint in the mass plane. The shaded region is excluded in the
lower-right panel of Fig.~\ref{1}.\footnote{Contrary to PSR
J1141$-$6545, the parameter $s$ is measured for PSR J0737$-$3039, and
the corresponding solid curve should not be considered as the mere
boundary of the shaded region. Its width is larger than $1\sigma$ error
bars.} But the greatest novelty of this system is that the companion
has also been observed as a pulsar.\cite{0737b} Of course, this second
pulsar cannot be recycled itself (since only the first formed could
accrete matter from its companion), therefore its pulse period is slow
($2.8$ s) and its observation is not precise enough to provide extra
post-Keplerian observables. However, its \textit{Keplerian} orbital
data do suffice to give an extra test. Indeed, if $x_A$ and $x_B$
denote the observed projected semi-major axis of both pulsars, the
ratio $x_A/x_B = m_B/m_A$ gives a very precise measure of the mass
ratio, independently of the theory (and valid at any order in powers of
$1/c$, at least in general relativity and tensor-scalar theories).
Moreover, if one exchanges the roles of bodies $A$ and $B$, the mass
function (\ref{4}) and the bound $|\sin i|\leq 1$ give a second
excluded region in the mass plane. This is the reason why the white
region has got a V shape in the lower-right panel of Fig.~\ref{fig1},
and why the shaded (excluded) region covers almost all the plane. This
will be quite important to constrain alternative theories of gravity in
Sec.~4 below. The orbital period derivative $\dot P$ of this
double-pulsar system has not yet been determined, but it should be
available with reasonable precision within a few months. It will of
course provide an extra strip in the mass plane, and thereby an extra
test of relativistic gravity in the strong-field regime. Geodetic
precession has already been observed in PSRs B1913+16 and B1534+12.
This confirms the existence of such an effect, and it can be used to
map the emission beam of the pulsar (which gives us information about
its structure), but this does not provide an actual test of the theory.
On the other hand, the geodetic precession period is predicted to be of
about 70 years for both pulsars in PSR J0737$-$3039, within general
relativity. Several years of observation should thus provide an extra
test of the theory, which would not have been possible with the other
known binary pulsars presented above. Let us finally mention that
because this double-pulsar system is almost perfectly edge on, one
observes eclipses of the recycled pulsar by its companion, and
modulations of the companion's pulses caused by the energy flux of the
recycled pulsar. Therefore, this system will allow us to probe pulsar
magnetospheres. Again, this cannot be considered as a test of the
gravity theory, but as a powerful new observatory of the physics of
neutron stars.

Besides the above four precisely timed binary pulsars, other systems do
provide extra tests, even if one does not measure enough post-Keplerian
parameters to determine accurately the two masses as in
Fig.~\ref{fig1}. Indeed, one can use a statistical argument on the
pulsar's mass $m_A \approx 1.35 m_\odot$ and the \textit{a priori}
arbitrary inclination angle $i$ to predict a probable value for the
companion's mass $m_B$ thanks to the Keplerian ``mass function''
(\ref{4}). Then, a single post-Keplerian observable may be compared to
the prediction of a theory. For instance, in the neutron star-white
dwarf binary \textbf{PSR B0655+64}, only an upper bound has been
obtained on the orbital period derivative $\dot P$. It is an order of
magnitude larger than the prediction of general relativity, therefore
this theory is obviously consistent again with experimental data.
However, this upper bound suffices to rule out a wide range of
alternative theories of gravity, which generically predict a value for
$\dot P$ several orders of magnitude larger (see Secs.~3 and 4 below).
Alternative theories also predict various effects which vanish
identically in general relativity, caused by violations of the strong
equivalence principle, of local Lorentz invariance or of conservation
laws. The experimental upper bounds on such effects therefore just
constitute ``null tests'' for general relativity, but have the
capability of constraining other theories. We will mention such an
example in Sec.~4 below.

\section{Tensor-scalar theories of gravity}
As shown is the previous section, several tests of gravity are
available in the strong-field regime, and general relativity passes
all of them with flying colors. We now wish to embed Einstein's theory
into a class of alternatives, in order to understand better which
features have been tested, and to compare the probing power of the
various tests. A generalization of the PPN formalism to all orders in
$1/c^n$ would need an infinite number of parameters. It is much more
efficient to restrict our study to the most natural class of
alternatives to general relativity, namely ``tensor-scalar'' theories,
in which gravity is mediated by a tensor field ($g_{\mu\nu}$) together
with one or several scalar fields ($\varphi$). This class of models is
privileged for many reasons. (i)~They are mathematically consistent
field theories, and do not involve any negative-energy mode nor any
adynamical field. (ii)~The existence of scalar partners to the graviton
is predicted by all unified and extra-dimensional theories, notably
superstrings. (iii)~Scalar fields play a crucial role in modern
cosmology, notably to explain the accelerated expansion phases of the
universe (inflation, quintessence). (iv)~Tensor-scalar models are the
only consistent massless field theories able to satisfy the weak
equivalence principle (universality of free fall of laboratory-size
objects). (v)~They are the only known theories satisfying ``extended
Lorentz invariance'', \textit{i.e.}, such that the gravitational
physics of subsystems, influenced by external masses, exhibit Lorentz
invariance. (vi)~They explain the key role played by
$\beta^\text{PPN}$ and $\gamma^\text{PPN}$ in the PPN formalism
(the extra 8 parameters introduced by Nordtvedt and Will\cite{w93}
vanish identically in tensor-scalar theories). (vii)~They are general
enough to describe many different deviations from general relativity,
but simple enough for their predictions to be fully worked
out.\cite{def1}

In this paper, we will further restrict our study to theories which
involve a single scalar field, and which satisfy exactly the weak
equivalence principle. Like in general relativity, the action of matter
is given by a functional $S_m[\psi_m, \widetilde g_{\mu\nu}]$ of some
matter fields $\psi_m$ (including gauge bosons) and one second-rank
symmetric tensor $\widetilde g_{\mu\nu}$. The difference with general
relativity lies in the kinetic term of $\widetilde g_{\mu\nu}$. Instead
of being a pure spin-2 field, it is here a mixing of spin-2 and spin-0
excitations. More precisely, it can be written as $\widetilde
g_{\mu\nu} = \exp [2a(\varphi)] g_{\mu\nu}$, where $a(\varphi)$ is a
function of a scalar field $\varphi$, and $g_{\mu\nu}$ is the Einstein
(spin 2) metric. The action of the theory reads thus
\begin{equation}
S = \frac{c^3}{16\pi G}\int d^4 x\sqrt{-g}\left(
R - 2g^{\mu\nu}\partial_\mu\varphi\partial_\nu\varphi\right)
+ S_m\left[\psi_m, e^{2a(\varphi)}g_{\mu\nu}\right].
\label{5}
\end{equation}
[Our signature is $\scriptstyle -+++$, $R$ is the scalar curvature of
$g_{\mu\nu}$, and $g$ its determinant.]

Our discussion will now be focused on the function $a(\varphi)$, which
characterizes the coupling of matter to the scalar field. Let us expand
it around the background value of the scalar field, which can be chosen
to vanish without loss of generality:
\begin{equation}
a(\varphi) = \alpha_0\varphi +\frac{1}{2}\beta_0
\varphi^2 +\cdots
\label{6}
\end{equation}
The slope $\alpha_0$ measures the coupling strength of the linear
interaction between matter and the scalar field, $\beta_0$
is the quadratic coupling constant of matter to two scalar lines,
\textit{etc.} [A diagrammatic representation is given to label the axes
of Figs.~\ref{fig4} and \ref{fig5} below.] General relativity
corresponds to a vanishing function $a(\varphi) = 0$, and
Jordan-Fierz-Brans-Dicke theory to a linear function $a(\varphi)=
\alpha_0 \varphi$, with $\alpha_0^2 = 1/(2\omega_\text{BD} +3)$. As
shown below, interesting strong-field effects occur when $\beta_0\neq
0$, \textit{i.e.}, when $a(\varphi)$ has a nonvanishing curvature.

At the first post-Newtonian order (\textit{i.e.}, when measuring
effects of order $1/c^2$ in weak-field conditions), the predictions
of tensor-scalar theories depend only on the first two parameters
$\alpha_0$ and $\beta_0$. The effective gravitational constant between
two bodies and the Eddington PPN parameters read:
\begin{subequations}
\label{7}
\begin{eqnarray}
G^\text{eff} & = & G(1+\alpha_0^2)\,,
\label{7a}\\
\gamma^\text{PPN}-1 & = & -2 \alpha_0^2/(1+\alpha_0^2)\,,
\label{7b}\\
\beta^\text{PPN}-1 & = & \frac{1}{2}\,
\frac{\alpha_0\beta_0\alpha_0}{(1+\alpha_0^2)^2}\,.
\label{7c}
\end{eqnarray}
\end{subequations}
[The factor $\alpha_0^2$ comes from the exchange of a scalar particle
between two bodies, whereas $\alpha_0\beta_0\alpha_0$ comes from a
scalar exchange between three bodies.] The solar-system bounds
(\ref{2})-(\ref{3}) therefore impose that both $\alpha_0^2$ and
$\alpha_0^2|\beta_0|$ must be small. This implies that the scalar field
must be linearly \textit{weakly} coupled to matter. On the other hand,
the quadratic coupling strength $\beta_0$ is not directly constrained
if $\alpha_0^2$ is small enough, and its sign can also be arbitrary.
[Note that $a(\varphi)$ is a coupling function and not a potential for
the scalar field, therefore a negative $\beta_0$ does not spoil the
stability of the field theory.]

\begin{figure}[ht]
\centerline{\epsfbox{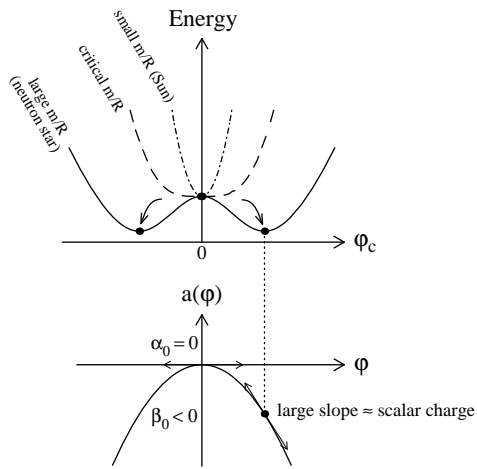}}
\caption{Heuristic argument to explain the phenomenon of ``spontaneous
scalarization''. When $\beta_0 <0$ and the compactness $Gm/Rc^2$ of a
body is large enough, it is energetically favorable to create a local
scalar field different from the background value. The body becomes
thus strongly coupled to the scalar field.
\label{fig2}}
\end{figure}

At higher post-Newtonian orders $1/c^n$, a simple diagrammatic
argument shows that any deviation from general relativity
involves at least two factors $\alpha_0$, and has the schematic form
\begin{equation}
\text{deviation from G.R.} = \alpha_0^2\times
\left[\lambda_0 + \lambda_1 \frac{Gm}{Rc^2} + \lambda_2
\left(\frac{Gm}{Rc^2}\right)^2+\cdots \right],
\label{8}
\end{equation}
where $\lambda_0$, $\lambda_1$, \dots\ are constants built from the
coefficients $\alpha_0$, $\beta_0$, \dots\ of expansion (\ref{6}).
Since $\alpha_0^2$ is experimentally known to be small, we thus expect
the theory to be close to general relativity at any order. However,
some nonperturbative effects may occur in strong-field conditions: If
the compactness $Gm/Rc^2$ of a body is greater than a critical value,
the square brackets of Eq.~(\ref{8}) can become large enough to
compensate even a vanishingly small $\alpha_0^2$. To illustrate this,
let us consider a model for which $\alpha_0$ vanishes strictly,
\textit{i.e.}, which is perturbatively equivalent to general
relativity: There is strictly no deviation from general relativity at
any order in a perturbative expansion in powers of $1/c$. A parabolic
coupling function $a(\varphi) = \frac{1}{2} \beta_0 \varphi^2$ suffices
for our purpose. At the center of a static body, the scalar field takes
a particular value $\varphi_c$, and it decreases as $1/r$ outside. The
energy of such a scalar field configuration involves two contributions,
coming respectively from the kinetic term and from the matter-scalar
coupling function in action (\ref{5}). As a rough estimate of its
value, one can write
\begin{equation}
\text{Energy} \approx \int\left[\frac{1}{2}(\partial_i\varphi)^2
+\rho\, e^{\beta_0\varphi^2/2} \right]
\approx  mc^2\left(
\frac{\varphi_c^2/2}{Gm/Rc^2}
+ e^{\beta_0\varphi_c^2/2}\right).
\label{9}
\end{equation}
When $\beta_0<0$, this is the sum of a parabola and a Gaussian, and if
the compactness $Gm/Rc^2$ is large enough, the function
$\text{Energy}(\varphi_c)$ has the shape of a Mexican hat, see
Fig.~\ref{fig2}. The value $\varphi_c =0$ now corresponds to a local
{\it maximum\/} of the energy. It is therefore energetically favorable
for the star to create a nonvanishing scalar field $\varphi_c$, and
thereby a nonvanishing ``scalar charge'' $a'(\varphi_c) =
\beta_0\varphi_c$. This phenomenon is analogous to the spontaneous
magnetization of ferromagnets.

\begin{figure}[ht]
\centerline{\epsfbox{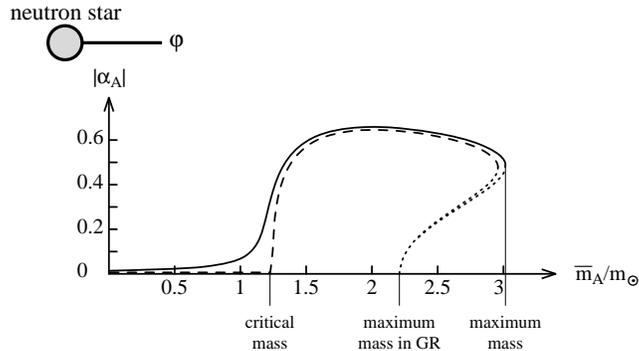}}
\caption{Scalar charge $\alpha_A$ versus baryonic mass $\overline m_A$,
for the model $a(\varphi) = -3\varphi^2$ (\textit{i.e.}, $\beta_0 =
-6$). The solid line corresponds to the maximum value of
$\alpha_0$ allowed by the Eqs.~(\ref{2}), and the dashed line to
$\alpha_0 = 0$. The dotted lines correspond to unstable configurations
of the star.
\label{fig3}}
\end{figure}

This heuristic argument has been verified by explicit numerical
calculations, taking into account the coupled differential equations of
the metric and the scalar field, and using various realistic equations
of state to describe nuclear matter inside a neutron
star.\cite{nonpert} The correct definition of the linear coupling
strength between a compact body $A$ and the scalar field reads
$\alpha_A\equiv \partial\ln m_A/\partial\varphi_0$. It is plotted in
Fig.~\ref{fig3} for the particular model $\beta_0 = -6$. One finds that
there exists indeed a ``spontaneous scalarization'' above a critical
mass (whose value decreases as $-\beta_0$ grows). On the other hand, if
$\beta_0 > 0$, both the above heuristic argument and the actual
numerical calculations show that $|\alpha_A| < |\alpha_0|$. In that
case, one finds that neutron stars are even less coupled to the scalar
field than solar-system bodies.

The scalar charge $\alpha_A$ enters the predictions of the theory
in the same way as $\alpha_0$ in weak-field conditions. For instance,
the effective gravitational constant between two bodies $A$ and $B$
reads
\begin{equation}
G_{AB}^\text{eff} = G (1+\alpha_A\alpha_B)\,,
\label{10}
\end{equation}
instead of Eq.~(\ref{7a}). Similarly, the strong-field analogues of the
Eddington parameters $\gamma^\text{PPN}$ and $\beta^\text{PPN}$
involve products of $\alpha_A$, $\alpha_B$,
$\beta_A\equiv\partial\alpha_A/\partial\varphi_0$ and
$\beta_B$, instead of $\alpha_0$ and $\beta_0$ as in Eqs.~(\ref{7}).
Since the scalar charge $\alpha_A \approx 0.6$ in the model of
Fig.~\ref{fig3}, one thus expects deviations by $\sim 35\%$ from some
general relativistic predictions. Moreover, the quadratic coupling
strength $\beta_A$ can take very large numerical values near the
critical mass, like the magnetic susceptibility of ferromagnets.
Therefore, even larger deviations from general relativity are found
when the mass of a neutron star happens to be close to the critical
one. An even more spectacular phenomenon occurs for the quantity
$\alpha_B \partial\ln I_A/\partial\varphi_0$, involved in the
post-Keplerian observable\footnote{This expression enters $\gamma_T$
because the inertia moment of the pulsar, $I_A$, is modified by the
presence of a companion at a varying distance.} $\gamma_T$. Indeed,
when $\beta_0<0$, this quantity blows up as $\alpha_0\rightarrow 0$:
Paradoxically, a theory which is closer to general relativity in
weak-field conditions predicts larger deviations in the strong-field
regime! On the other hand, when $\beta_0 > 0$, strong-field predictions
are even closer to those of general relativity than in the solar
system.

The post-Keplerian observable which is the most affected by the
presence of a scalar partner to the graviton is the orbital period
derivative $\dot P$. Indeed, the energy flux carried out by
gravitational waves is of the form
\begin{eqnarray}
&&\text{Energy flux} = \left\{\frac{\text{Quadrupole}}{c^5} +
\mathcal{O}\left(\frac{1}{c^7}\right)\right\}_\text{helicity 2}
\nonumber\\
&&+ \left\{\frac{\text{Monopole}}{c} +
\frac{\text{Dipole}}{c^3} + \frac{\text{Quadrupole}}{c^5} +
\mathcal{O}\left(\frac{1}{c^7}\right)\right\}_\text{helicity 0}.
\label{11}
\end{eqnarray}
The first curly brackets contain the prediction of general relativity.
The second ones contain the extra contributions predicted in
tensor-scalar theories. The powers of $1/c$ give the orders of
magnitude of the different terms. In particular, the monopolar and
dipolar helicity-0 waves are generically expected to be much larger
that the usual quadrupole of general relativity. However, the scalar
monopole has the form
\begin{equation}
\frac{\text{Monopole}}{c} =
\frac{G}{c}\left\{\frac{\partial(m_A\alpha_A)}{\partial t}
+\frac{\partial(m_B\alpha_B)}{\partial t} +
\mathcal{O}\left(\frac{1}{c^2}\right)\right\}^2,
\label{12}
\end{equation}
and it reduces to order $\mathcal{O}(1/c^5)$ if the stars $A$ and $B$
are at equilibrium, $\partial_t(m_A\alpha_A) = 0$, which is the case
for all binary pulsars quoted in Sec.~2 above. However, this monopolar
term would be huge in the case of a collapsing star, for instance. The
dipole has the form
\begin{equation}
\frac{\text{Dipole}}{c^3} = \frac{G}{3c^3}
\left(\frac{G_{AB}m_Am_B}{r_{AB}^2}\right)^2 (\alpha_A-\alpha_B)^2 +
\mathcal{O}\left(\frac{1}{c^5}\right),
\label{13}
\end{equation}
and is usually much larger that a quadrupole of order $1/c^5$. For
instance, in a pulsar-white dwarf binary, the pulsar's scalar charge
$\alpha_A$ may be of order unity, like in Fig.~\ref{fig3} above,
whereas the weakly self-gravitating white dwarf has a very small scalar
charge $\alpha_B \approx \alpha_0$, constrained by Eqs.~(\ref{3}) and
(\ref{7b}). On the other hand, in a double-neutron star system, one
expects $m_A \approx m_B$ and therefore $\alpha_A \approx \alpha_B$, so
that this dipolar contribution is considerably reduced (but may still
be large with respect to the usual quadrupole of general relativity).
Indeed, a dipole is a vector in space, and two strictly identical stars
do not define a preferred orientation.

It should be noted that even within Brans-Dicke theory, \textit{i.e.},
for a linear coupling function $a(\varphi) = \alpha_0 \varphi$, the
dipolar contribution (\ref{13}) does not vanish identically. Indeed, in
that case, one can show that $\alpha_A = \alpha_0 (1- 2 s_A)$,
where $s_A \approx Gm/Rc^2$ is the compactness of the star, so that
$(\alpha_A-\alpha_B)^2 = 4\alpha_0^2(s_A-s_B)^2$. The dipolar
contribution is thus proportional to the experimentally small factor
$\alpha_0^2 = 1/(2\omega_\text{BD} +3)$, but it does not vanish, and
can still be larger than the usual quadrupole if $\alpha_0^2 >
v^2/c^2 \sim 10^{-6}$. Note also that because $s_A \approx Gm/Rc^2$,
this dipole (\ref{13}) is formally of order $\mathcal{O}(G^3/c^7)$, so
that one could not have obtained it by a naive calculation at linear
order in $G$. It is however crucial to take it into account, because it
is numerically of order $\mathcal{O}(G/c^3)$ since $Gm/Rc^2 \approx
0.2$ for a neutron star.

\section{Experimental constraints on tensor-scalar theories}
We saw in the previous section that solar-system experiments probe
only the first two coefficients $\alpha_0$ and $\beta_0$ of the
matter-scalar coupling function (\ref{6}). On the other hand, the
strong-field predictions do depend on the full shape of this function.
In order to compare easily the different constraints, we will
restrict our study to a strictly parabolic function $a(\varphi) =
\alpha_0\varphi + \frac{1}{2}\beta_0\varphi^2$.
\begin{figure}[ht]
\centerline{\epsfbox{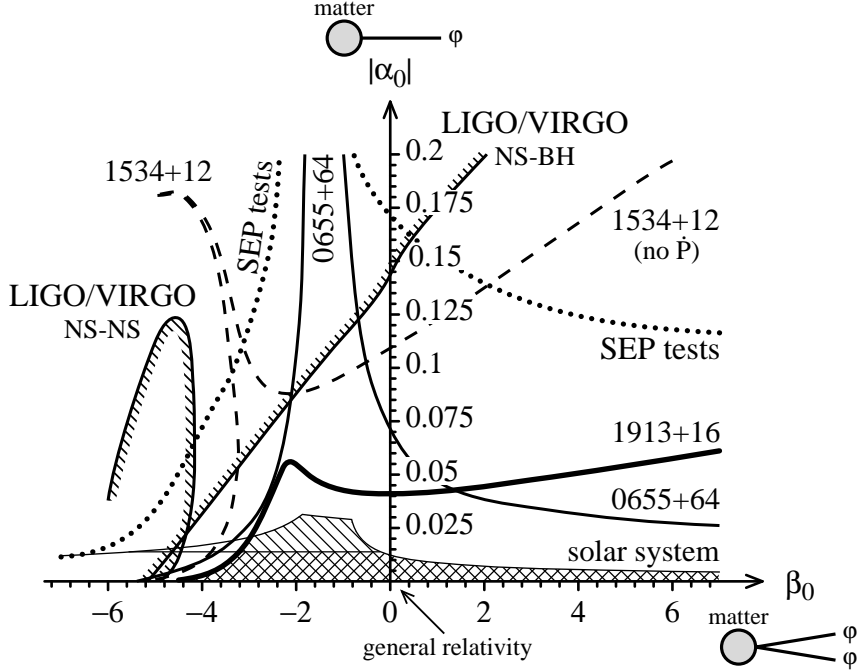}}
\caption{Constraints on generic tensor-scalar theories imposed by
solar-system experiments, classic binary-pulsar tests, and future
detections of inspiralling binaries with laser interferometers. The
hatched region is allowed by all the tests. The doubly hatched one is
also consistent with VLBI data.$^4$\label{fig4}}
\end{figure}
Figure~\ref{fig4} displays the known constraints\cite{nonpert} at the
time of the previous Marcel Grossmann meeting MG9, three years ago.
Since the physics does not depends on the sign of $\alpha_0$, only a
half plane $(|\alpha_0|, \beta_0)$ is drawn. Each point on this figure
represents a tensor-scalar theory. General relativity is at the origin
$\alpha_0=\beta_0=0$, Brans-Dicke theory is on the vertical axis
$\beta_0=0$, and the horizontal axis $\alpha_0=0$ corresponds to
theories which are perturbatively equivalent to general relativity (see
Sec.~3 above). For each theory $(|\alpha_0|, \beta_0)$, the mass planes
of the various binary pulsars can be plotted like in Fig.~\ref{fig1},
and if the different strips do not have a common intersection, the
theory is ruled out. The allowed theories lie under and to the right of
the various curves.

The thin solid line labeled ``solar system'' corresponds to the bounds
(\ref{2}), and the unpublished limit provided by very long baseline
interferometry\cite{eubanks} (VLBI) is materialized by a thin
horizontal line. As discussed in the previous section, they impose a
very small matter-scalar linear coupling strength
$\alpha_0$, but do not constrain the quadratic coupling $\beta_0$ if
$\alpha_0$ is small enough.

On the other hand, binary pulsars impose $\beta_0 > -4.5$, even for a
vanishingly small $\alpha_0$. As discussed in the previous section,
this constraint is due to the spontaneous scalarization of neutron
stars, which occurs when $-\beta_0$ is large enough.
Equations~(\ref{7}) allow us to rewrite this inequality in terms of the
Eddington parameters $\beta^\text{PPN}$ and $\gamma^\text{PPN}$,
which are both consistent with 1 in the solar system. One finds
\begin{equation}
\frac{\beta^\text{PPN}-1}{\gamma^\text{PPN}-1} < 1.1\,.
\end{equation}
The singular ($0/0$) nature of this ratio vividly expresses why such a
conclusion could net be obtained in weak-field experiments, and
underlines that binary-pulsar tests are qualitatively different.

This bound is mainly due to the Hulse-Taylor binary pulsar \textbf{PSR
B1913+16}. For \textbf{PSR B1534+12}, whose orbital period derivative
$\dot P$ is not very well known, we have plotted the constraints
imposed only by the four post-Keplerian observables $\dot\omega$,
$\gamma_T$, $r$ and
$s$. This gives much weaker constraints than PSR B1913+16, but this is
nevertheless a very important test because it does not depend on the
radiative structure of the theory. Taking into account the measured
value of $\dot P$ removes the horn-shaped region at the top-left of
the dashed line.

The second most constraining system is the neutron-star white dwarf
\textbf{PSR B0655+64}, although it has not been timed precisely enough
to determine the two component masses, and that only an upper
experimental bound on its orbital period derivative $\dot P$ is known.
Indeed, even by choosing conservative values for the two masses, such
an asymmetrical system loses too much energy through dipolar scalar
waves in most scalar-tensor theories, see Eq.~(\ref{13}).

Equation (\ref{10}) above tells us that the acceleration of a neutron
star $A$ towards the center $C$ of the Galaxy is proportional to
$(1+\alpha_A\alpha_C)$, whereas a white dwarf $B$ is accelerated with a
force $\propto (1+\alpha_B\alpha_C)$. Since $\alpha_A\neq \alpha_B$ in
general, there is a violation of the strong equivalence principle
(\textbf{SEP}). This causes a polarization of the orbit of a neutron
star-white dwarf system towards the Galaxy center, analogous to the
Stark effect in electromagnetism. More precisely, the eccentricity
vector $\mathbf{e}$ of the orbit is the sum of a fixed vector
$\mathbf{e}_F$ directed towards the Galaxy center (proportional to the
difference of the accelerations of the two bodies) and of a rotating
vector $\mathbf{e}_R(t)$ corresponding to the usual relativistic
periastron advance at angular velocity $\dot \omega_R$. Several
asymmetric systems of this kind (such as PSRs 1713+0747, 2229+2643,
1455$-$3330) happen to have a very small eccentricity. The only
explanation would be that the rotating vector $\mathbf{e}_R(t)$ is
precisely canceling the fixed contribution $\mathbf{e}_F$ at the time
of our observation: $\mathbf{e}_F + \mathbf{e}_R(t) \approx
\mathbf{0}$. However, this is very improbable if the system is old
enough, and one can use a statistical argument to constrain the space
of theories. Moreover, by considering several such systems, the
probability that they have simultaneously a small eccentricity is the
product of the already small individual probabilities. The allowed
tensor-scalar theories lie between the two (approximate) dotted lines.
This test is much less constraining than the others because the Galaxy
is not compact enough to be spontaneous scalarized, therefore its
scalar charge $\alpha_C\approx \alpha_0$ is small, and the difference
of the accelerations $\propto (\alpha_A-\alpha_B)\alpha_C \approx
(\alpha_A-\alpha_0)\alpha_0$ is small too.

To detect the gravitational wave signal from an inspiralling binary
with the \textbf{LIGO/VIRGO} interferometers, one will perform a
matched filter analysis of the signal, using the gravitational wave
templates predicted by general relativity. In tensor-scalar theories,
the waveforms are very different, because of the extra contributions of
the scalar waves to the energy loss, Eq.~(\ref{11}). Therefore, the
actual detection of an inspiralling binary with the general
relativistic templates will constrain the magnitude of the scalar
contributions. The hatched curve labeled ``LIGO/VIRGO NS-BH'' displays
the region of the theory space which would be excluded if a $1.4
m_\odot$ neutron star--$10 m_\odot$ black hole system is detected with
a signal to noise ratio $S/N = 10$ (the excluded region lie on the
hatched side). The detection of a double neutron star system with
masses similar to those of PSR B1913+16 would exclude the bubble of
theories labeled ``LIGO/VIRGO NS-NS''. As we can see on
Fig.~\ref{fig4}, these regions are \textit{already} excluded by
binary-pulsar tests. Therefore, we can conclude that the general
relativistic wave template do suffice for these interferometers: Even
if there exists a scalar partner to the graviton, we anyway already
know that it is too weakly coupled to matter to change significantly
the waveforms. This is a good news, because the inclusion of possible
scalar contributions would have considerably slowered the data
analysis. On the other hand, it has been proven\cite{lisa} that the
LISA space interferometer could still be sensitive to scalar-field
effects. Indeed, the detection of a neutron star inspiralling a $1000
m_\odot$ black hole would probe values of $|\gamma^\text{PPN}-1|$ as
small as $4\times 10^{-6}$, \textit{i.e.}, an order of magnitude
tighter than the best present bounds (\ref{3}) [or a factor 3 tighter
for the matter-scalar coupling strength $\alpha_0$]. However, we will
see below that the binary pulsar PSR J1141$-$6545 will probably probe
similar values of $|\gamma^\text{PPN}-1|$ around 2010. Therefore, it
may not be necessary to start including scalar corrections to the wave
templates for LISA: This binary pulsar should tell us, just before the
launch of the LISA mission, whether there is or not a scalar
contribution at this order.

\begin{figure}[ht]
\centerline{\epsfbox{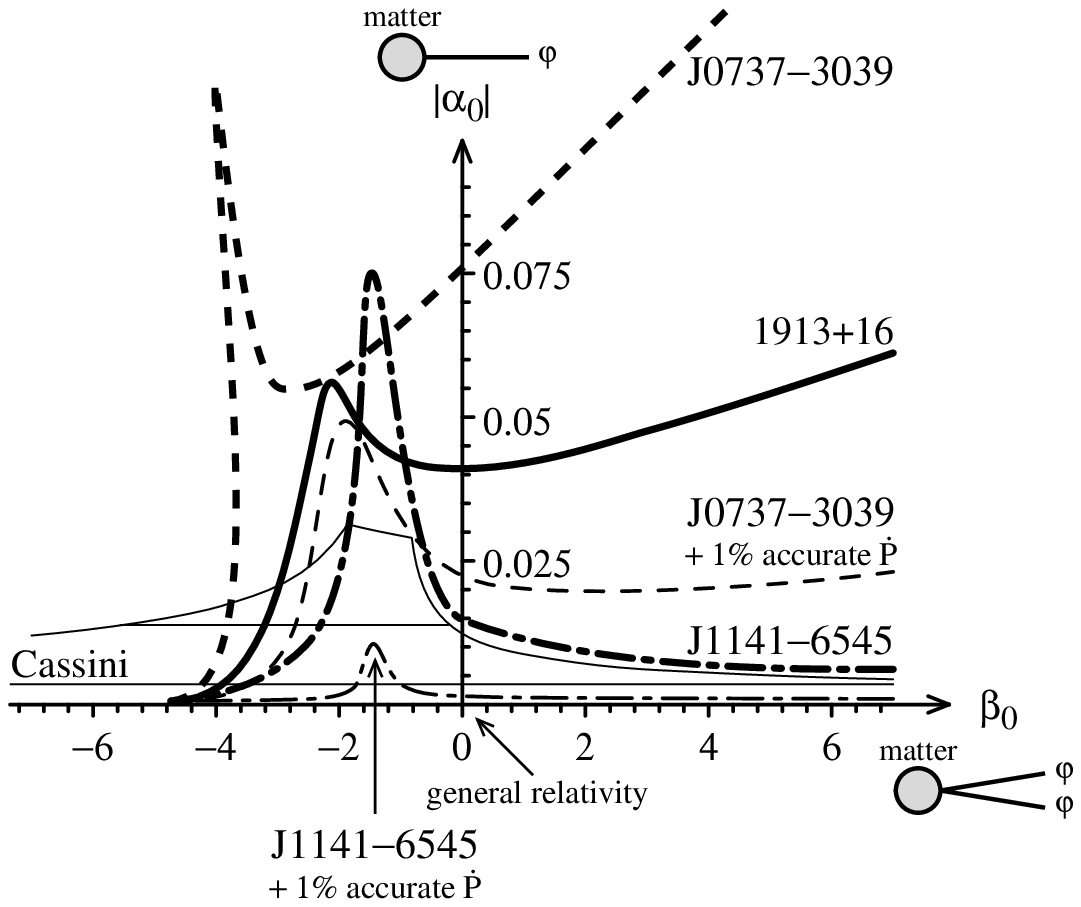}}
\caption{Constraints imposed on the theory space by the two recently
timed binary pulsars PSR J1141$-$6545 and J0737$-$3039. The thinner
lines display the constraints which will be reached when the orbital
period derivative $\dot P$ is determined with $1\%$ accuracy. We also
display here the impressive solar-system constraint (\ref{3}) obtained
recently from the observation of the Cassini spacecraft near solar
conjunction.\protect\cite{cassini}
\label{fig5}}
\end{figure}

The constraints imposed by the recently timed \textbf{PSR J1141$-$6545}
are much tighter than the previous ones, and we plot them as a
dot-dashed curve in Fig.~\ref{fig5}. We present here preliminary
results, which will be refined in a forthcoming
publication.\cite{def04} Note that the vertical scale of this Figure
has been expanded by a factor $2$ with respect to Fig.~\ref{fig4}. To
ease the comparison of these two Figures, we have repeated in
Fig.~\ref{fig5} the curves corresponding to PSR B1913+16 and to
solar-system constraints.
The reason why PSR J1141$-$6545 is so constraining is because this is
an asymmetrical system, composed of a neutron star and a white dwarf.
Therefore, like PSR B0655+64 above, it generically emits a large amount
of dipolar scalar waves, which are inconsistent with the small measured
value of its orbital period derivative $\dot P$. The lower-left panel
of Fig.~\ref{fig1} shows that the experimental precision on this
quantity is not very good yet. However, this is already much better
than the mere upper bound on $\dot P$ that we knew for PSR B0655+64.

Like all binary pulsars, PSR J1141$-$6545 excludes tensor-scalar
theories corresponding to $\beta_0 < -4.5$, even for a strictly
vanishing $\alpha_0$. This is again due to the phenomenon of
spontaneous scalarization described in Sec.~3. But it is quite
remarkable that this system is also very constraining in the region of
positive $\beta_0$'s. It is almost as powerful as the previously known
solar-system bounds (\ref{2}). This may seem paradoxical, because we
saw that when $\beta_0 > 0$, strongly self-gravitating bodies are
much more weakly coupled to the scalar field than matter in the solar
system. The reason why this binary pulsar anyway provides a powerful
test in this region of the theory plane is again its asymmetry. Indeed,
the pulsar's scalar charge is exponentially small, $|\alpha_A| \ll
|\alpha_0|$, whereas the weakly self-gravitating white dwarf companion
has a standard (background) scalar charge $\alpha_B \approx \alpha_0$.
Therefore, the dipolar radiation term (\ref{13}), proportional to
$(\alpha_A-\alpha_B)^2 \approx \alpha_0^2$, is non negligible, and the
small observed value of $\dot P$ is still constraining. The asymptotic
limit of the bound on $|\alpha_0|$ for $\beta_0 \rightarrow +\infty$
can also be estimated analytically,\footnote{Our formal limit
$\beta_0 \rightarrow +\infty$ actually means $\beta_0 \sim 10$ or $30$,
but for extremely large values of this parameter, the
``de-scalarization'' phenomenon would affect the white dwarf too, and
the constraints would thus weaken.} by imposing that the dipolar
contribution (\ref{13}) is smaller than the experimental error
$\Delta \dot P$. One finds
\begin{equation}
\alpha_0^2 < \frac{96}{5} \left(\frac{2 \pi G(m_A+m_B)}{P
c^3}\right)^{2/3}
\frac{1+\frac{73}{24}e^2+\frac{37}{96}e^4}{(1+\frac{e^2}{2})(1-e^2)}
\left|\frac{\Delta \dot P}{\dot P}\right|
\approx 5\times 10^{-5}\left|\frac{\Delta \dot P}{\dot P}\right|.
\end{equation}
With the present uncertainty on $\dot P^\text{obs} = (-4\pm 1)\times
10^{-13}$, this gives $|\alpha_0| < 0.004$ for $\beta_0 \rightarrow
+\infty$, consistently with the curve plotted in Fig.~\ref{fig5}.
As mentioned is Sec.~2, the precision on $\dot P$ should reach $1\%$ by
the end of the decade. This would give an asymptotic bound
$|\alpha_0| < 7\times 10^{-4}$ for $\beta_0 \rightarrow +\infty$,
corresponding to $|\gamma^\text{PPN}| < 10^{-6}$, more than one order
of magnitude tighter than the recent limit (\ref{3}). This estimate is
confirmed by our numerical plot, in Fig.~\ref{fig5}, of the constraints
that PSR J1141$-$6545 will impose when this precision is reached.
Of course, a detailed analysis of the possible sources of noise will be
necessary then, notably of the tidal effects on the white dwarf, and
of the Doppler contribution due to the acceleration of the system
towards the center of the Galaxy.

\begin{figure}[ht]
\centerline{\epsfbox{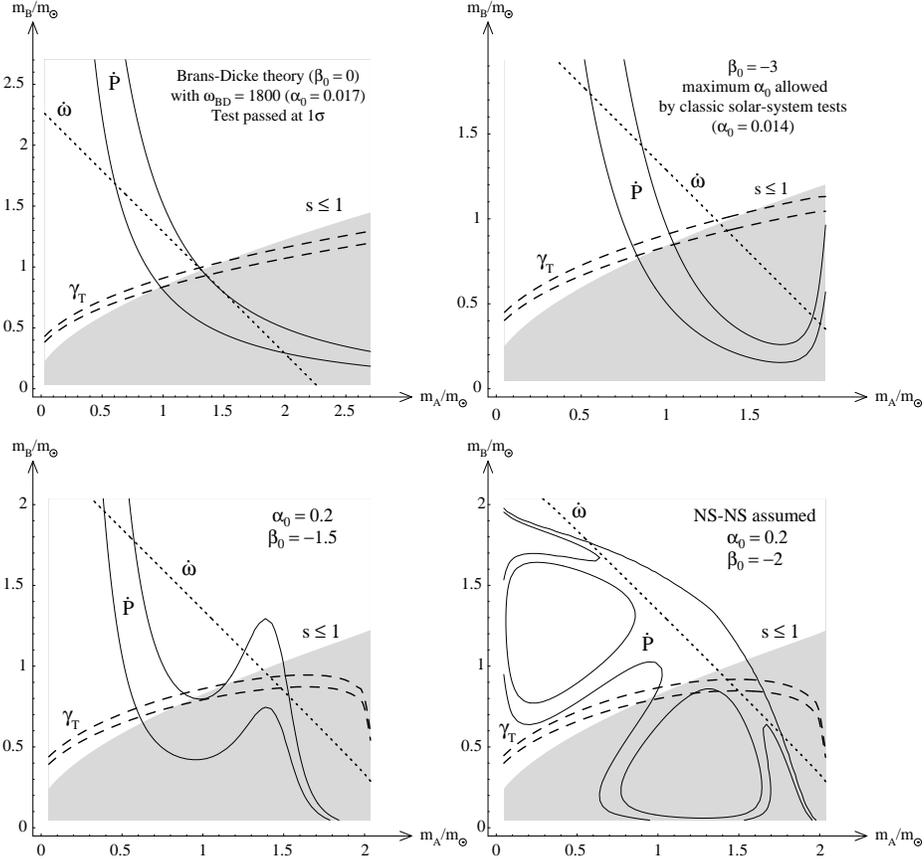}}
\caption{Mass plane ($m_A=$ pulsar, $m_B$ = companion) for PSR
J1141$-$6545 in four different tensor-scalar theories of gravity.
Only the upper-left plot corresponds to an allowed model (at the
$1\sigma$ level). In the lower-right plot, the system is assumed to be
a double-neutron star system.
\label{fig6}}
\end{figure}

To give the reader a better feeling of the above limits, we display in
Fig.~\ref{fig6} the mass plane for PSR J1141$-$6545 in four different
scalar-tensor theories.
The upper-left plot corresponds to Brans-Dicke theory, \textit{i.e.},
to a linear matter-scalar coupling function $a(\varphi) =
\alpha_0\varphi$, with $\alpha_0^2 = 1/(2\omega_\text{BD}+3)$. The
value of $\alpha_0$ has been fine tuned to pass the test exactly at the
$1\sigma$ level. This theory corresponds thus to the intersection of
the thick dot-dashed curve of Fig.~\ref{fig5} with the vertical axis.
The corresponding limit $|\gamma^\text{PPN}|< 6\times 10^{-4}$
($\Leftrightarrow \omega_\text{BD} > 1800$) is not far from the
unpublished VLBI constraint,\cite{eubanks} as can be also directly seen
on Fig.~\ref{fig5}. Note that in Fig.~\ref{fig6}, the $\dot P$
strip has been significantly displaced with respect to its location
in general relativity (lower-left panel of Fig.~\ref{fig1}).

The upper-right panel of Fig.~\ref{fig6} displays the mass plane for a
theory which was allowed by \textit{all} experimental data at the time
of the MGX conference, but which is violently ruled out (by $5\sigma$)
by PSR J1141$-$6545. Indeed, the three strips do not have any common
intersection. This illustrates that this binary pulsar is much more
constraining than the others, at least in the privileged class of
tensor-scalar theories. However, the subsequent solar-system bound
(\ref{3}) does even better.

The lower-left panel of Fig.~\ref{fig6} provides an example of a theory
in which the three strips $\gamma_T$, $\dot\omega$ and $\dot P$ do have
a common intersection, but it is located is the shaded region
corresponding to $|\sin i| > 1$ (see Sec.~2 above). [The lower-right
panel gives another example.] Therefore, in such a case, this
(Keplerian) mathematical constraint becomes crucial. If the inequality
$|\sin i| \leq 1$ is not taken into account, the bump of the thick
dot-dashed curve blows up to very large values of $|\alpha_0|$, in
Fig.~\ref{fig5}. Note however that such theories are anyway ruled out
by solar-system experiments.

The lower-right panel of Fig.~\ref{fig6} illustrates the strange shapes
that the curves can take in some theories. Here, the topology of the
$\dot P$ ``strip'' has even changed. Note also that the intersection of
the strips corresponds to masses which are significantly different from
those obtained in general relativity (lower-left panel of
Fig.~\ref{fig1}). Therefore, it would be inconsistent to use the masses
obtained within general relativity to compute the predicted $\dot P$ in
another theory. Obviously, all the strips must be computed within the
\textit{same} theory, and the binary-pulsar test is passed if they have
a common intersection, even if its location differs from that obtained
in general relativity. Contrary to the other three mass planes, the
lower-right one has been computed while assuming that PSR J1141$-$6545
is a double-neutron star system. This is excluded at the $90\%$ level
by the formation scenario\cite{1141form} of this system. However, it
remains instructive to study the dependence of the constraints on the
nature of the companion. If it were a neutron star, the system would be
much more symmetric ($\alpha_A\approx\alpha_B$), therefore the dipolar
contribution (\ref{13}) to the energy flux would be much lower, and the
observed value of $\dot P$ less constraining. We did plot the
corresponding bounds in the theory plane $(|\alpha_0|, \beta_0)$, but
we do not display them here to clarify the Figures. It suffices to
mention than PSR J1141$-$6545 would give constraints similar to those
of PSR B0655+64 if it were a double neutron star system (see
Fig.~\ref{fig4}).

Let us come back to Figure~\ref{fig5} above, where we also plotted as
dashed lines the bounds imposed by the double pulsar \textbf{PSR
J0737$-$3039}, at present and when a $1\%$ accurate $\dot P$ is
available. Curiously enough, in spite of the great precision of the
measures, and although the mass ratio $m_A/m_B = 1.07$ is tightly fixed
independently of the theory, this system appears not to be extremely
constraining for scalar-tensor theories. When a $1\%$ accurate $\dot P$
is measured, it will of course provide a much stronger test, and become
more constraining than the Hulse-Taylor binary pulsar PSR B1913+16.
However, it pales in comparison with the neutron star-white dwarf
binary PSR J1141$-$6545, even with its present large uncertainties on
$\dot P$. The reason is that the two components of PSR J0737$-$3039 are
neutron stars with similar scalar charges, and therefore that its
dipolar radiation (\ref{13}) is weak. One may also wonder why at
present, without any observed $\dot P$, the constraints imposed by this
system are so loose, as compared to those of PSR B1913+16. The reason
seems to be that the five observed functions of the masses
(\textit{cf.} the lower-left panel of Fig.~\ref{1}) depend only weakly
on the presence of a scalar field. Two of them, $r$ and $x_A/x_B$, are
even totally insensitive to it. Only $\gamma_T$ can vary a lot when
spontaneous scalarization occurs (because of the blowing contribution
$\alpha_B \partial\ln I_A/\partial\varphi_0$, see Sec.~3), but this
post-Keplerian parameter happens to have still large experimental
uncertainties.

\begin{figure}[ht]
\centerline{\epsfbox{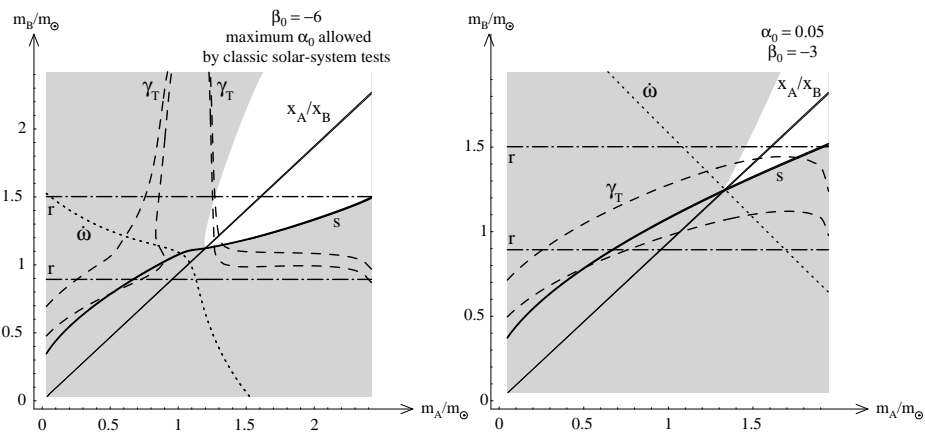}}
\caption{Mass plane ($m_A=$ pulsar, $m_B$ = companion) for PSR
J0737$-$3039 in two different tensor-scalar theories of gravity.
The left plot corresponds to a theory which is ruled out by all
binary-pulsar tests, including the present one. The theory considered
in the right panel passes the present test, although it is ruled out by
several other binary pulsars.
\label{fig7}}
\end{figure}

Figure~\ref{fig7} displays the mass plane $(m_A,m_B)$ for PSR
J0737$-$3039 within two tensor-scalar theories.
The left panel illustrates that this system has the capability of
ruling out some models. As shown in Fig.~\ref{5}, it forbids all
theories with $\beta_0 < -4.5$, even for a strictly vanishing
$\alpha_0$, \textit{i.e.}, even if the theory is strictly
indistinguishable from general relativity in the weak-field conditions
of the solar system. This property is shared by all binary pulsars.
They exhibit similar deformations of the various strips in the
mass plane, as soon as $-\beta_0$ is large enough. Notice in particular
the characteristic shape of the $\gamma_T$ strip, caused by the
spontaneous scalarization of neutron stars.

On the contrary, the right panel of Fig.~\ref{fig7} illustrates that
this system is presently less constraining than several other binary
pulsars. Indeed, the model $\alpha_0 = 0.05$, $\beta_0 = -3$ is
inconsistent with PSRs B1913+16, B0655+64, and J1141$-$6545, but it
passes the test for the double pulsar J0737$-$3039. This mass plane
confirms that the five observables are weakly dependent on the scalar
field. The largest deformation occurs for the $\gamma_T$ strip, with
respect to general relativity (lower-right panel of Fig.~\ref{fig1}),
but its width is large enough for the test to be passed. Of course,
such a model will be also ruled out by this double pulsar J0737$-$3039
once its $\dot P$ is measured with reasonable precision.

Let us recall that this double pulsar will anyway provide brand new
tests of relativistic gravity, as well as very important information
about the astrophysics of pulsars, although its discriminating power
seems rather weak at present in the framework of tensor-scalar
theories. Moreover, if the new tests happen not to be passed by general
relativity, one can already bet that no tensor-scalar theory will be
able to pass them either. Therefore, this system has actually the
capability of ruling out the best class of gravity theories as a whole.
On the other hand, if general relativity passes the new tests, as we
expect, this double pulsar will not teach us much about tensor-scalar
models.

\section{Conclusions}
Binary pulsars are ideal tools for testing relativistic gravity in the
strong field regime. In the most natural class of alternatives to
general relativity, tensor-scalar theories, their probing power has
been shown to be qualitatively different from weak-field experiments:
They have the capability of testing theories which are strictly
equivalent to general relativity in the solar system.

Two fantastic binary pulsars have been timed recently. The double
pulsar J0737$-$3039 promises to be the best laboratory for testing
general relativity itself, and for studying the physics of pulsars.
On the other hand, the neutron star-white dwarf system PSR J1141$-$6545
is by far the most constraining binary pulsar known at present, because
its asymmetry implies that it generically emits strong dipolar
gravitational waves in scalar-tensor theories. It is already almost as
constraining as solar-system tests even in the region of positive
$\beta_0$'s, where binary pulsars never competed with weak-field
experiments up to now. It should probe values of the Eddington
parameter $|\gamma^\text{PPN}-1|\sim 10^{-6}$ by the end of the decade,
\textit{i.e.}, more than one order of magnitude better than present
solar-system limits.

Binary pulsars are so precise that they already exclude the models
which would have predicted some significant scalar-field contributions
to the gravitational wave templates necessary for LIGO and VIRGO.
Therefore, one may use securely the general relativistic templates for
these interferometers. It is still possible that such scalar-field
effects be detectable with the LISA space interferometer, but binary
pulsars will probably give us tighter bounds before it is launched.

\end{document}